\documentclass[aps,prd,amsmath,onecolumn,notitlepage,showpacs,superscriptaddress,nofootinbib,usenatbib,11pt]{revtex4-1}
\def \be {\begin{equation}} 
\def \ee {\end{equation}} 
\def \bea {\begin{eqnarray}} 
\def \eea {\end{eqnarray}} 

\usepackage{graphicx}
\usepackage{dcolumn}
\usepackage{bm}
\usepackage{epsfig} 
\usepackage{amsfonts}
\usepackage{amsmath}
\usepackage{amssymb}
\usepackage[usenames]{color}
\usepackage[dvipsnames]{xcolor}
\usepackage[unicode, colorlinks=true, linkcolor=linkcolor, citecolor=linkcolor, filecolor=linkcolor,urlcolor=linkcolor, pdfusetitle]{hyperref}

\hypersetup{colorlinks,citecolor=blue,linkcolor=blue,urlcolor=blue}
\hypersetup{final=true}

\newcommand*{\Om}{$\mathcal{O}_{\rm m}$~}
\newcommand*{\Lm}{$\mathcal{L}_{\rm m}$~}

\begin{document}

\title{Null tests of the concordance model in the era of Euclid and the SKA}

\author{Carlos A. P. Bengaly}
\email{carlosbengaly.on.br}
\affiliation{Observat\'orio Nacional, 20921-400, Rio de Janeiro - RJ, Brazil}
\affiliation{D\'epartement de Physique Th\'eorique, Universit\'e de Gen\`eve, 1211 Gen\'eve 4, Switzerland}
\affiliation{Department of Physics \& Astronomy, University of the Western Cape, Cape Town 7535, South Africa}

\author{Chris Clarkson}
\email{chris.clarkson@qmul.ac.uk}
\affiliation{School of Physics \& Astronomy, Queen Mary, University of London, United Kingdom}
\affiliation{Department of Mathematics \& Applied Mathematics, University of Cape Town, Cape Town  7701, South Africa}
\affiliation{Department of Physics \& Astronomy, University of the Western Cape, Cape Town 7535, South Africa}

\author{Martin Kunz}
\email{Martin.Kunz@unige.ch}
\affiliation{D\'epartement de Physique Th\'eorique, Universit\'e de Gen\`eve, 24 quai Ernest Ansermet, 1211 Gen\'eve 4, Switzerland}

\author{Roy Maartens}
\email{roy.maartens@gmail.com}
\affiliation{Department of Physics \& Astronomy, University of the Western Cape, Cape Town 7535, South Africa}
\affiliation{Institute of Cosmology \& Gravitation, University of Portsmouth, Portsmouth PO1 3FX, United Kingdom}

\date{\today}

\begin{abstract}
We perform null tests of the concordance model, using $H(z)$ measurements that mimic next-generation surveys such as Euclid and the SKA. To this end, we deploy a non-parametric method, so that we make minimal assumptions about the fiducial cosmology as well as the statistical analysis. We produce simulations assuming different cosmological models in order to verify how well we can distinguish between their signatures. We find that SKA- and Euclid-like surveys should be able to discriminate sharply between the concordance and alternative dark energy models that are compatible with the Planck CMB data. We conclude that SKA and Euclid will be able to falsify the concordance model in a statistically significant way, if one of the benchmarks models represents the true Universe, without making assumptions about the underlying cosmology. 
\end{abstract}

\maketitle

\newpage

\section{Introduction}\label{sec:intro}

The flat $\Lambda$CDM cosmological model has been established as the concordance model (CM) of Cosmology in the past two decades. It is based upon the fundamental assumptions of (a)~the Cosmological Principle -- so that cosmic distances and ages are described by the Friedmann-Lema\^itre-Robertson-Walker (FLRW) metric -- and (b) General Relativity as the theory of gravity, and it incorporates an accelerated phase over the last few billion years. Recent measurements from the cosmic microwave background (CMB)~\cite{planck18}, Type Ia Supernova distances (SNIa)~\cite{scolnic18}, and large-scale structure clustering (LSS)~\cite{icaza-lizaola19, des19}, are able to constrain its cosmological parameters to very good precision. Despite its success in accommodating all current cosmological observations, the CM faces problems, both theoretical (e.g. the vacuum energy problem) and observational (e.g. conflicting measurements of the Hubble constant~\cite{verde19, riess20}). These problems motivate the development of further observational tests to probe the consistency and the foundations of the CM.   

Next-generation redshift surveys like Euclid~\cite{euclid13, euclid18, euclid19}, SKA~\cite{ska20} and DESI~\cite{desi16} will deliver measurements of the Hubble parameter from baryonic acoustic oscillations (BAO) with unprecedented precision, facilitating  sub-percent precision on key cosmological parameters. In this work, we use the upcoming precision as a probe of the CM itself. Rather than a parametric analysis, i.e., model-fitting of alternative cosmological models in order to quantify deviations from a flat $\Lambda$CDM model, we carry out null tests based on general consistency relations that the CM must obey. These consistency relations are formulated in terms of functions of $H(z)$ and its derivatives, which are constant or zero if the Universe is described by $\Lambda$CDM \emph{regardless} of the parameters of the model. In this way, we can determine how well we will be able rule out the CM \emph{without} prior assumptions on the underlying cosmological parameters. Such a non-parametric analysis avoids biasing results by fitting specific cosmological models, so that our analysis is {\em independent} of cosmological assumptions. Rather, our results only rely on the choice of kernel reconstruction, which is shown to be robust regardless of this choice.\footnote{Note that $H(z)$ measurements from BAO depend on the fiducial cosmology via the sound horizon $r_s$ at the drag epoch. However, these measurements are calibrated with respect to $\Lambda$CDM to ensure that this assumption is consistent with observations. Measurements of cosmological distances, like luminosity distances of standard candles and  sirens, as well as angular diameter distances from  BAO, are also model-independent in a similar sense -- but our null tests would rely on the second derivative of these measurements, which would significantly degrade the results.}

\newpage

\section{Method}\label{sec:method}

Our non-parametric approach is based on Gaussian processes, which are distributions over functions, rather than over variables as in the case of standard Gaussian distributions over parameters. We can thus reconstruct a function from data points without assuming a parametrisation, a method which is robust for interpolation as well as extrapolation~\cite{link1}. We deploy the code {\sc GaPP} (Gaussian Processes in Python)~\citep{seikel12} (see also~\citep{shafieloo12}) to reconstruct $H(z)$ and $dH/dz$ from  simulated data-sets. Similar methods and applications have been used previously~\cite{shafieloo10, yahya13, sapone14, busti14, costa15, gonzalez16, joudaki17, yu18, marra18, gomezvalent18a, haridasu18, gomezvalent18b, keeley20, bengaly20a, bengaly20b, arjona20, benisty20, mukherjee20a, mukherjee20b,briffa20, vonMarttens20, colgain21, perenon21, escamilla-rivera21, bernardo21}, but not for the forecasts that we develop here. Note that that we only optimise the $H(z)$ reconstruction, but not its derivative.

The Hubble parameter $H(z)$ in a generic dark energy model is given by
\begin{eqnarray}\label{eq:hz}
w(z) &=& p_{\rm DE}(z)/\rho_{\rm DE}(z)\,; \\
E(z)^2 &\equiv& \left[\frac{H(z)}{H_0}\right]^2 = \Omega_{\rm m}(1+z)^3 + 
(1-\Omega_{\rm m}-\Omega_{\rm de})(1+z)^2 + \Omega_{\rm de} \exp{\left[3\int_0^z \frac{1+w(\tilde z)}{1+\tilde z}d\tilde z\right]} \;.
\end{eqnarray}
We assume a flat $\Lambda$CDM cosmology as in the CM:
\begin{eqnarray}\label{eq:omega_DE}
\Omega_{\rm de} = 1-\Omega_{\rm m} \,, \qquad w(z)  = -1 \,,
\end{eqnarray}
with fiducial parameter values given by Planck 2018 (TT, TE, EE+lowE+lensing) best-fits: 
\begin{eqnarray}\label{eq:model1}
H_0 = 67.36 \pm 0.54 \, \mathrm{km \, s}^{-1} \, \mathrm{Mpc}^{-1} \,, 
\qquad
\Omega_{\rm m} = 0.3166 \pm 0.0084 \,. 
\end{eqnarray}

The null test that we apply is based on the consistency relation for the CM model~\cite{zunckel08} (see also~\cite{sahni08, clarkson08} for similar tests):
\begin{equation}\label{eq:om}
\mathcal{O}_{\rm m}(z) \equiv \frac{E(z)^2-1}{(1+z)^3 - 1} = \Omega_{\rm m} \;\; \mbox{in FLRW}\,,
\end{equation}
which can be obtained from~\eqref{eq:hz} and \eqref{eq:omega_DE}. Then we have the null test: 
\begin{equation}\label{eq:null_test_01}
\mathcal{O}_{\rm m}(z) \neq \Omega_{\rm m} \;\; \mbox{implies concordance model ruled out.}
\end{equation}
Differentiating~\eqref{eq:om} with respect to the redshift, we find a related property of the CM:
\begin{equation}\label{eq:lm}
\mathcal{L}_{\rm m}(z) \equiv 3(1+z)^2\big[1-E(z)^2\big] + 2z(3 + 3z + z^2)E(z)E'(z)=0 \,,
\end{equation}
which leads to an alternative null test:
\begin{equation}\label{eq:null_test_02}
\mathcal{L}_{\rm m}(z)  \neq 0 \;\; \mbox{implies concordance model ruled out.}
\end{equation}

In summary, if $\mathcal{O}_{\rm m}(z)-\Omega_{\rm m}$ differs from zero at a statistically significant level, then flat $\Lambda$CDM is ruled out. Similarly if $\mathcal{L}_{\rm m}(z) $ differs from zero at a statistically significant level. The second test~\eqref{eq:null_test_02} involves the computation of the first derivative of $E(z)$ data. Although this degrades the results, it is more effective to measure deviations from zero than from a constant value -- and in addition, we do not know the true $\Omega_{\rm m}$ value. 

In order to evaluate the  performance of future data-sets, we simulate $H(z)$ measurements in three different cosmological models:
\begin{eqnarray} \label{models}
\mbox{$K\Lambda$CDM:~~} & & \Omega_{\rm K} \equiv 1-\Omega_{\rm m}-\Omega_{\rm de}=-0.01 \,, \\
\mbox{$(w_0,w_a)$CDM:~~} & & w(z) = w_0 + w_a(1-a)\,, \; \mbox{where}
\notag \\
\mbox{CPL1:~~} & & \{w_0,w_a\}=\{-1.1,-1.0\} \,,\\
\mbox{CPL2:~~} & & \{w_0,w_a\}=\{-0.8,-0.4\} \,.
\end{eqnarray}
These models break the consistency relations~\eqref{eq:om} and~\eqref{eq:lm}, but they are still possible within the bounds imposed by current CMB-only observations~\cite{planck18}. 

We assess the statistical significance of the $\mathcal{L}_{\rm m}(z)$ test using the parameter 
\begin{equation}\label{eq:fz}
f(z_i) = \left|{{\mathcal{L}_{\rm m}(z_i)\over \sigma_i}}\right| \,,
\end{equation}
where the index $i$ denotes each individual GP test-point used for the reconstruction for a total number of $n_{\rm pts}$. We assume $n_{\rm pts}=100$ as the default number of GP test-points for the \Lm reconstruction, unless stated otherwise. As we have a continuous range of values for $f(z)$ across the test-points, we quote the maximum value of $f(z)$, hereafter $f_{\rm max}$, obtained across the redshift range of the survey as a measure of the maximum departure between these models and the CM reference value that the data allows. Larger value for $f(z)$ indicates a model that can be more easily distinguished from $\Lambda$CDM, i.e., $\mathcal{L}_{\rm m}=0$ for all redshift ranges. This test will be referred as $f_{\rm max}$-test from now on\footnote{The $f_{\rm max}$ test corresponds to performing a single $\chi^2$ evaluation at the redshift where we obtain the strongest constraint.}. Note that we do not apply this estimator to the \Om test since we do not know the true $\Omega_{\rm m}$ value. Furthermore, we stress that we are only deploying the $f_{\rm max}$-test for the sake of evaluating the survey performance on ruling out the null condition~\ref{eq:null_test_02}. A full analysis using the \Lm parameter as a probe of dark energy models will be pursued in the future.   

We simulate $H(z)$ measurements using the specifications of 3 next-generation surveys:\\
\underline{SKA-like intensity mapping survey}~\cite{ska20}:\\
$~\qquad\qquad\qquad \mbox{Band 1:}   
\quad 0.35<z<3.06\,, ~ N=20\,, \qquad
\mbox{Band 2:~~}   
\quad  0.10~<z~<0.50\,, ~ N=10\,.$ \\
\underline{Euclid-like galaxy survey}~\cite{ska20, euclid18, euclid19}:  
 \qquad $0.90<z<1.80\,, ~ N=20\,.$ \\
\underline{DESI-like galaxy survey}~\cite{desi16}:  
\qquad\quad $0.65<z<1.85\,, ~ N=20\,.$

Here $N$ is the number of data points that we assume, evenly distributed across the redshift range. 
The relative uncertainties, $ \sigma_H(z)/H(z)$,  that we use are taken from the interpolated curves in Figure 10 (left) of~\cite{ska20} (see also~\cite{bull16, euclid13}) for SKA- and Euclid-like surveys, while for the DESI-like case, we interpolate from Table 2.3 of~\cite{desi16}. No correlations were assumed between these $H(z)$ measurements as the measurements are extracted from power spectra estimated in wide redshift bins. Those spectra are generally taken to be independent (e.g.\ \cite{euclid19}). We also produce simulations assuming 30\% smaller uncertainties for $H(z)$ measurements. Hence, we determine how we can improve the performance of these null tests in case of reduced systematics, or slightly more futuristic surveys following similar specifications.

\section{Results}\label{sec:res}

\begin{figure*}[!ht]
\includegraphics[width=0.49\textwidth]{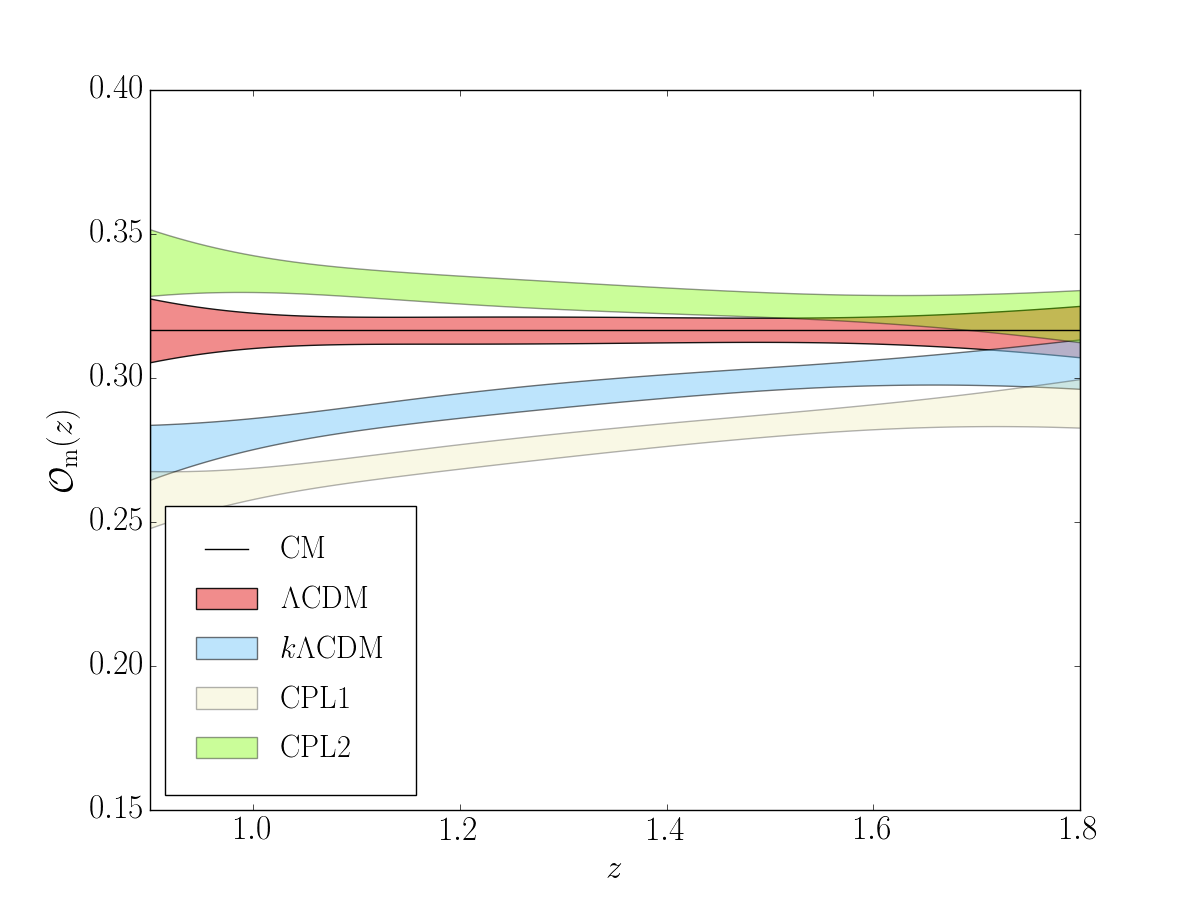}
\includegraphics[width=0.49\textwidth]{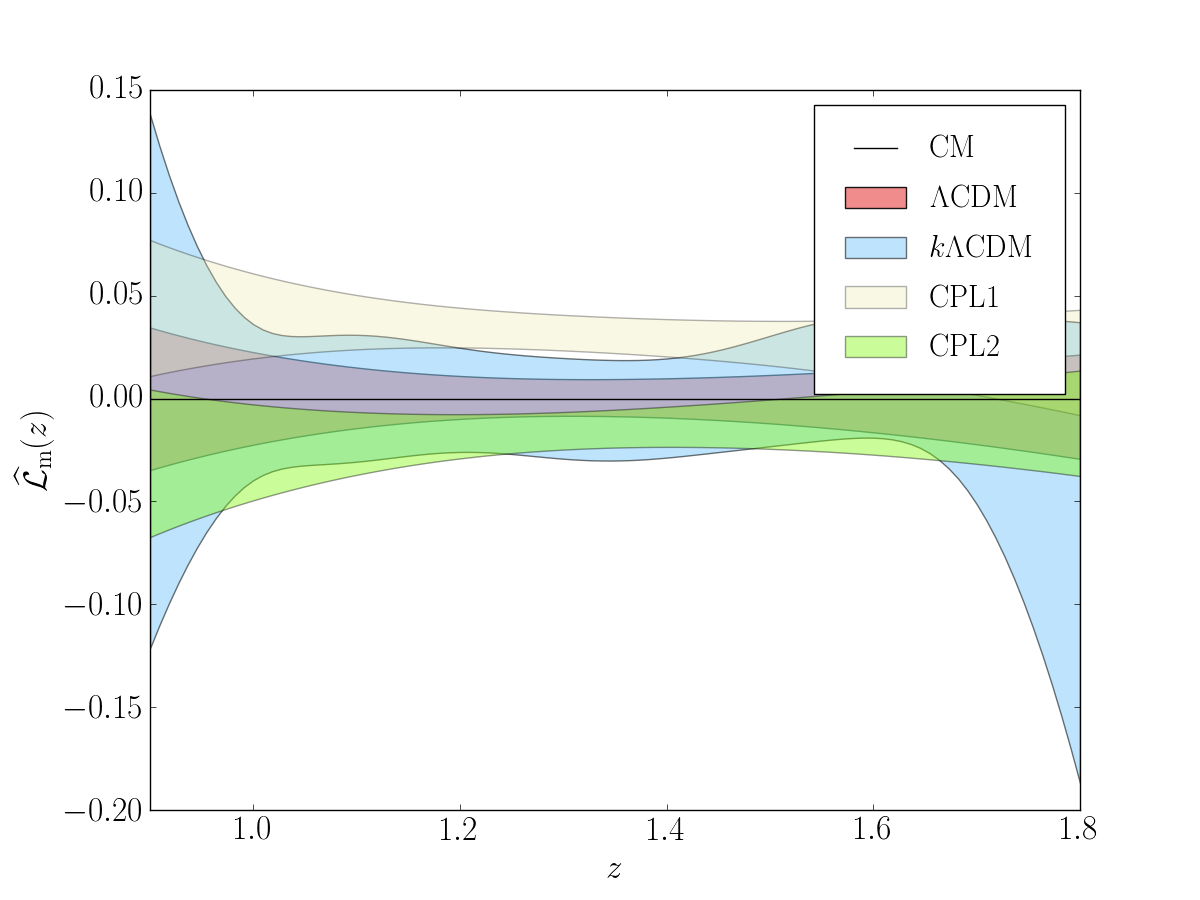}
\includegraphics[width=0.49\textwidth]{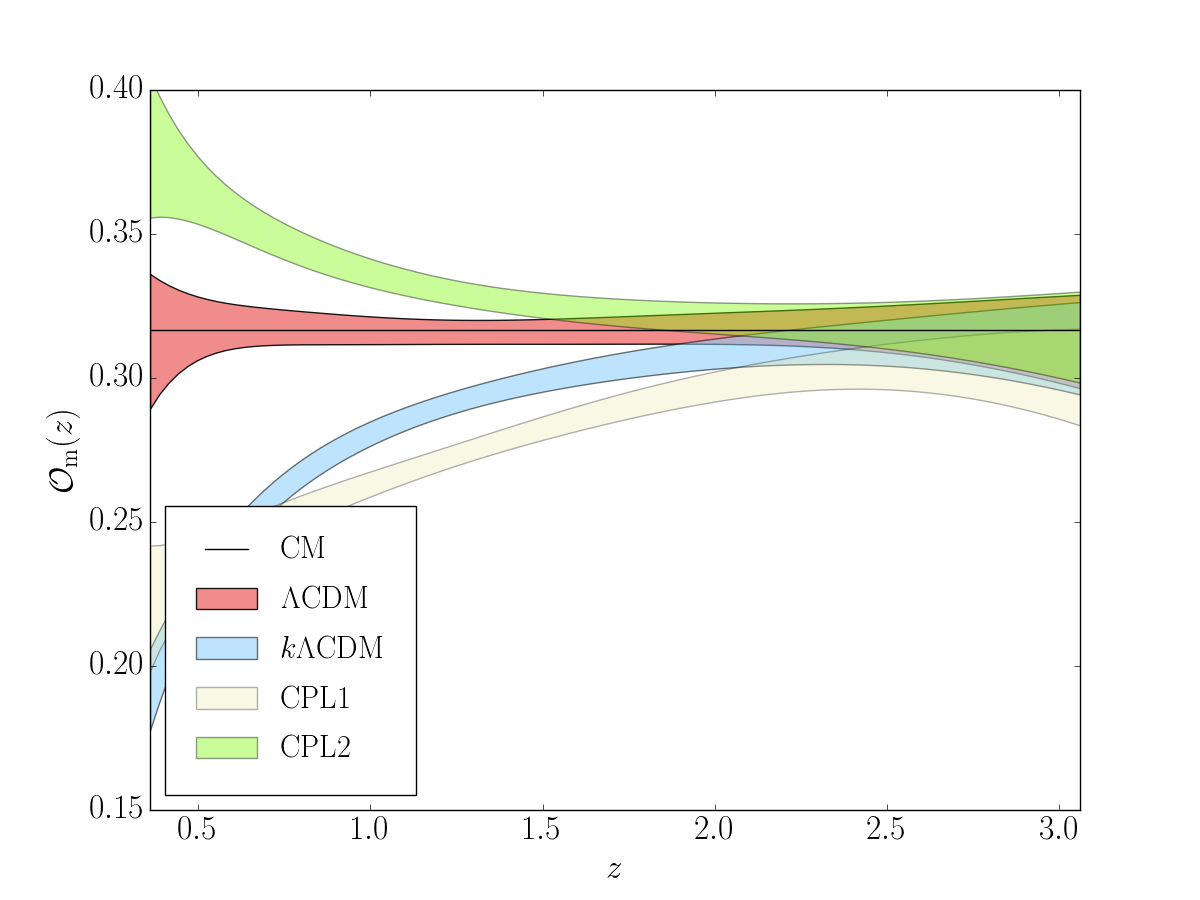}
\includegraphics[width=0.49\textwidth]{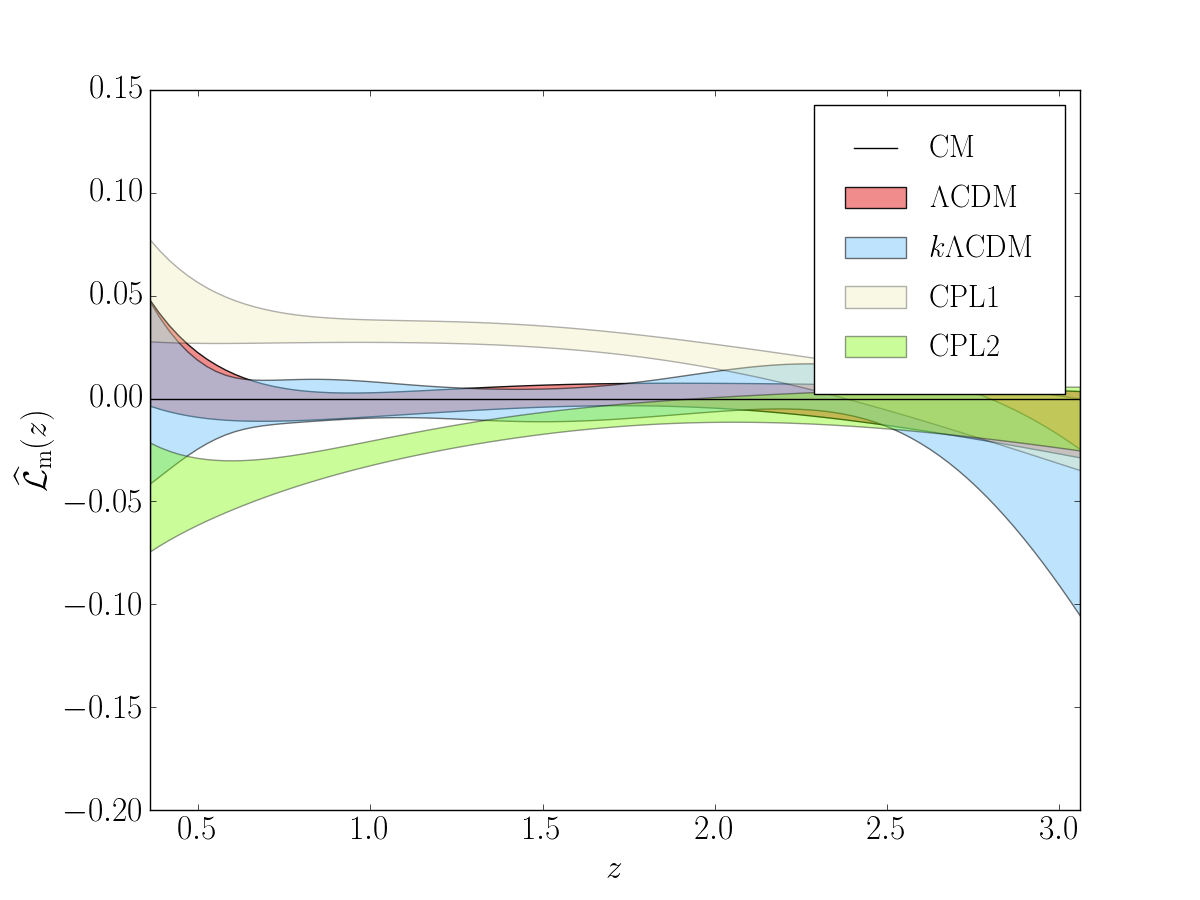}
\caption{\Om (left) and  \Lm (right) null tests, for Euclid-like (top) and SKA-like B1 (bottom) surveys. Shaded regions show $5\sigma$ ( \Om) and $3\sigma$ ( \Lm) CL for the reconstructed mean. 
$\widehat{\cal L}_{\rm m}(z) \equiv \mathcal{L}_{\rm m}(1+z)^{-6}$ is used rather than  \Lm to improve visualisation.}
\label{fig:rec_euclid_ska}
\end{figure*}

\begin{figure*}[!ht]
\includegraphics[width=0.49\textwidth]{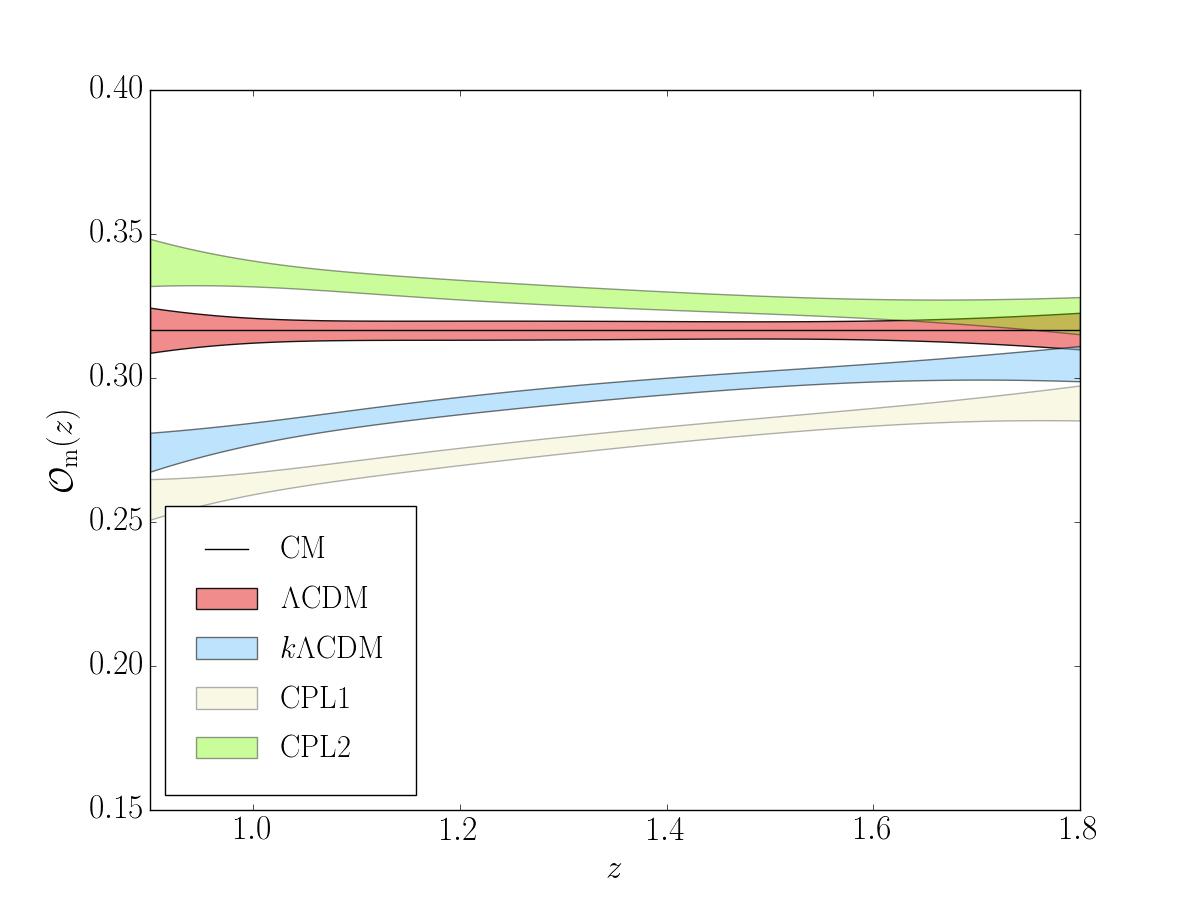}
\includegraphics[width=0.49\textwidth]{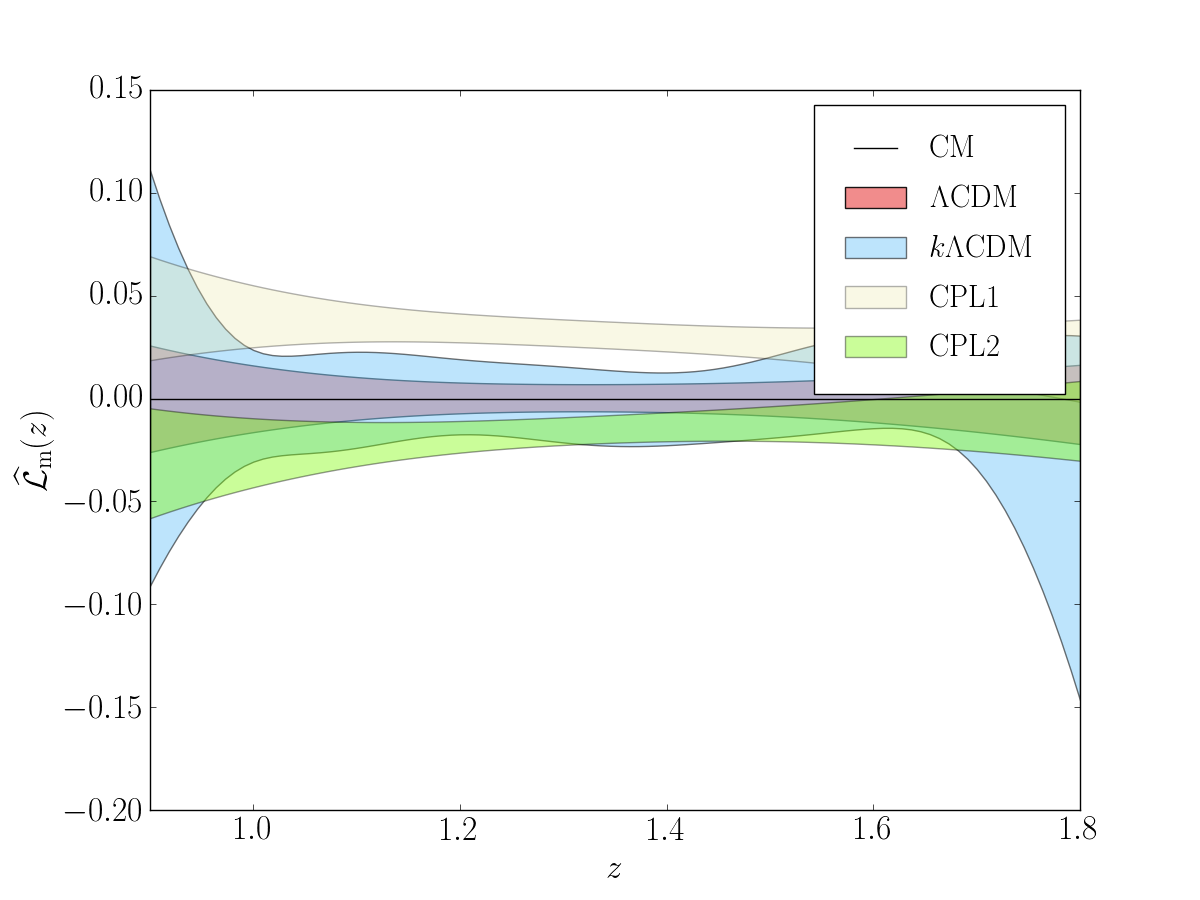}
\includegraphics[width=0.49\textwidth]{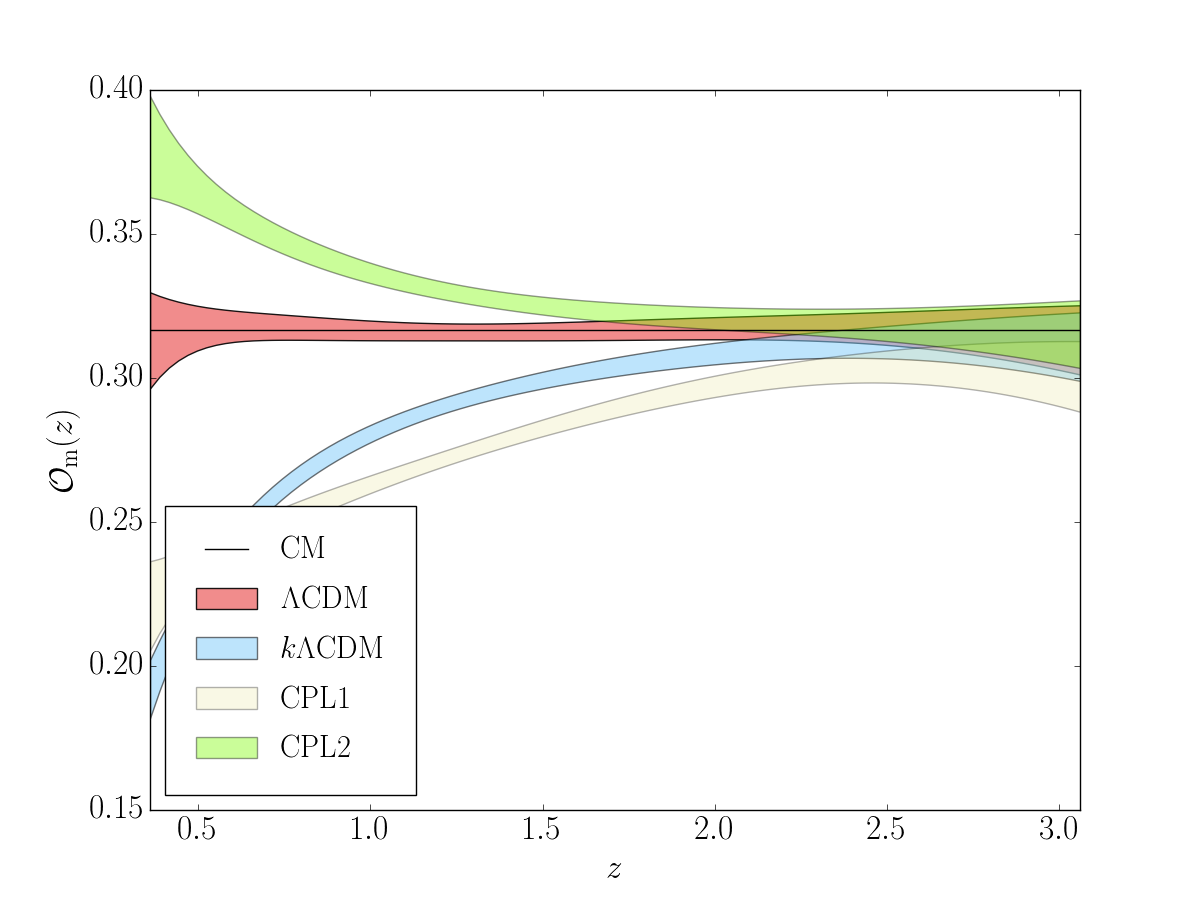}
\includegraphics[width=0.49\textwidth]{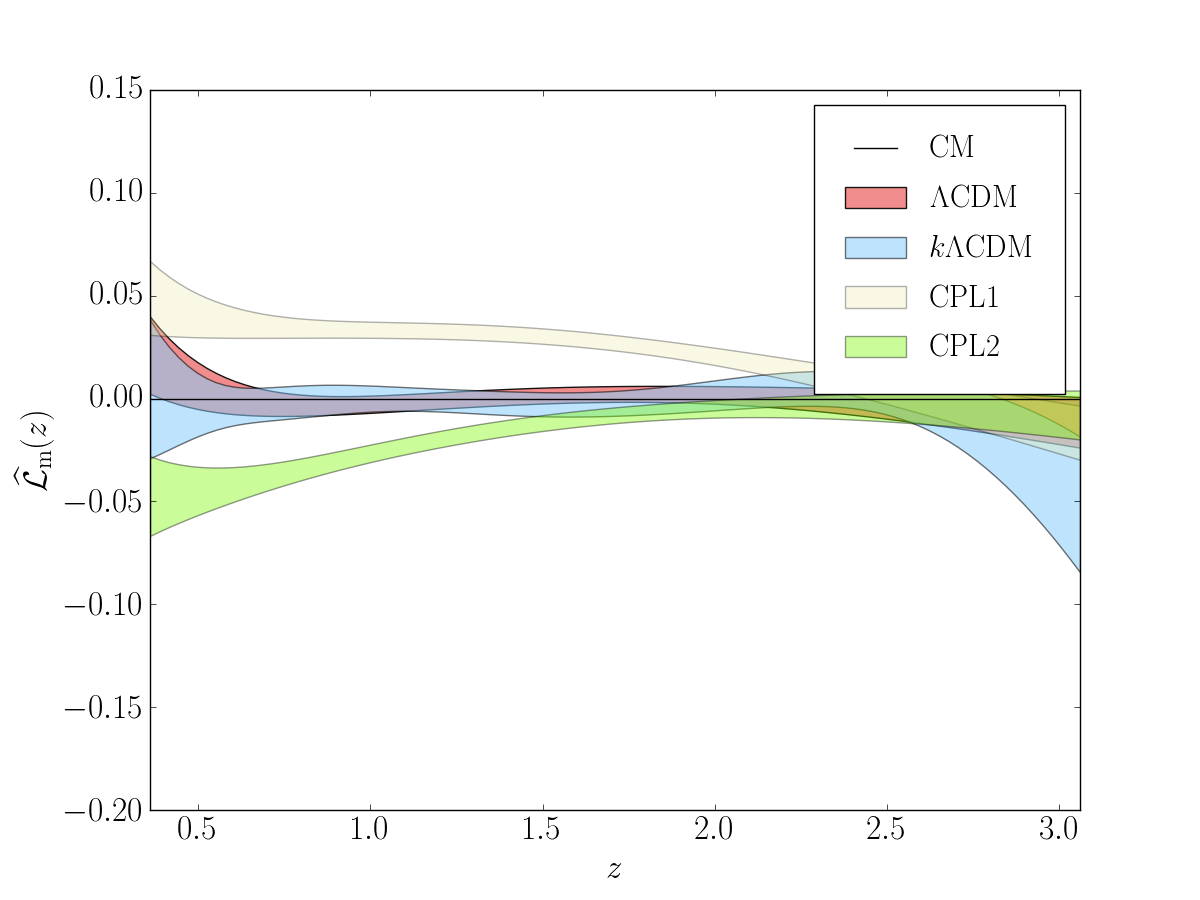}
\caption{Same as Fig.~\ref{fig:rec_euclid_ska}, but with $H(z)$ measurement uncertainties reduced by 30\%.}
\label{fig:rec_euclid_ska_0p7sig}
\end{figure*}

\begin{table*}[!t]
\begin{tabular}{ccccc}
\hline 
\hline 
\quad survey \quad \; & \quad \quad model \quad \quad \;\; & \quad $f_\mathrm{max}$ \; & $f_\mathrm{max} \; (0.7\sigma_H/H)$\\
\hline
\hline
	      & $\Lambda$CDM & $0.468$ & $0.445$ \\ 
Euclid-like   & $K\Lambda$CDM & $2.007$ & $1.964$ \\
	     & CPL1 & $11.228$ & $14.948$  \\
	     & CPL2 & $5.278$ & $7.340$   \\
\hline	     
		  & $\Lambda$CDM & $0.863$ & $1.125$  \\ 
SKA-like IM B1  &  $K\Lambda$CDM & $1.606$ & $1.559$  \\
                &  CPL1 & $6.136$ & $8.664$  \\
	        &  CPL2 & $4.674$ & $6.614$  \\
\hline	       
		&  $\Lambda$CDM & $0.586$ & $0.554$  \\ 
DESI-like      &  $K\Lambda$CDM & $2.448$ & $2.307$\\
		& CPL1 & $3.126$ & $4.538$  \\
	       &  CPL2 & $1.739$ & $2.377$  \\
\hline	       
		&  $\Lambda$CDM & $1.384$ & $1.707$  \\ 
SKA-like IM B2 	& $K\Lambda$CDM & $4.431$ & $4.395$  \\
		&  CPL1 & $1.400$ & $1.878$  \\
		& CPL2 & $1.820$ & $2.286$  \\
\hline
\hline
\end{tabular}
\caption{$f_{\rm max}$-test results for each survey and model, using the initial and optimistic uncertainties.} 
\label{tab:chi2} 
\end{table*}

We show the results for the \Om and \Lm tests using Euclid- and SKA B1-like simulations in Fig.~\ref{fig:rec_euclid_ska}. We plot $\widehat{\cal L}_{\rm m} \equiv \mathcal{L}_{\rm m}(1+z)^{-6}$ rather than the original \Lm to enhance the visualisation of the results, as is done in~\cite{seikel12, yahya13}. It is apparent that both surveys can distinguish between $\Lambda$CDM and other models at over $5\sigma$ confidence level (CL) for the \Om test, and at over $2\sigma$ CL for the \Lm case -- except for the highest redshift ranges reached by the SKA-like measurements, since their expected uncertainties are larger. The CPL1 simulations exhibit the largest departure from the CM, implying that future measurements can comfortably distinguish a dynamical dark energy model with these features,  {\em without} any prior assumption about dark energy itself. 

In Fig.~\ref{fig:rec_euclid_ska_0p7sig}, we show the results  for a more optimistic scenario with 30\% smaller uncertainties in $H(z)$ measurements. We see that improvements in the performance of the \Lm test mean that is could rule out the CM at over $3\sigma$ CL.

The statistical significance of the \Lm test results is presented in Table~\ref{tab:chi2}. We find that the $f_\mathrm{max}$ results for Euclid- and SKA B1-like surveys for $K\Lambda$CDM only mildly deviate from the $\Lambda$CDM case: $f_\mathrm{max} \simeq 2.0$ and $f_\mathrm{max} \simeq 1.6$ for Euclid- and SKA B1-like surveys, respectively. By contrast, the CPL models exhibit significant departure -- especially the CPL1 model, which gives $f_\mathrm{max} \simeq 11.2$ for a Euclid-like survey, and $f_\mathrm{max} \simeq 6.1$ for a SKA B1-like survey. These figures are lower for a DESI-like survey ($f_\mathrm{max} \simeq 3.1$ for CPL1, for instance), while a SKA-B2 like survey fails to rule out CM at a statistically significant level. Simulations with reduced uncertainties provide larger $f_\mathrm{max}$ for CPL models for both Euclid- and SKA B1-like surveys.  
 
We check the robustness of these results as follows. We repeat the analysis assuming different GP kernels, namely Mat\'ern(9/2) and Mat\'ern(7/2), obtaining for Euclid-like simulations: $f_\mathrm{max} \simeq 2.0$ and $f_\mathrm{max} \simeq 1.8$ for $K\Lambda$CDM; $f_\mathrm{max} \simeq 10.3 \; (5.0)$ and $f_\mathrm{max} \simeq 9.7 \; (4.8)$ for CPL1 (CPL2). Similar values were obtained for other survey simulations, and also when we use $n_{\rm pts}=500$ and $n_{\rm pts}=1000$ rather than $100$. 

\section{Conclusions}\label{sec:conc}

We applied null tests designed to rule out the concordance (flat $\Lambda$CDM) model, using simulated data for next-generation surveys. The  $\mathcal{O}_{\rm m}(z)$ and $\mathcal{L}_{\rm m}(z)$ null tests are based on consistency relations that only hold true if the Universe is described by the concordance model. 

We simulated $H(z)$ uncertainties from BAO measurements, using specifications and forecasts for Euclid-, SKA- and DESI-like spectroscopic surveys, and applied a non-parametric Gaussian process analysis to interpolate through these data. For a qualitative understanding of the discriminating power of these surveys, we used three  models different from flat $\Lambda$CDM but still compatible with Planck (CMB-only) 2018 constraints: a closed model, $K\Lambda$CDM, and two dynamical dark energy models following the CPL parametrisation. We also simulated $H(z)$ measurements with 30\% smaller uncertainties, so we can quantify how these tests improve in case of more controlled systematics, or in case of future surveys following similar specifications.

We found that Euclid- and SKA-like band 1 surveys can distinguish between $\Lambda$CDM and the alternative dark energy models, specially if we can reduce $H(z)$ uncertainties by $\sim$30\%. This was quantified through the $f_{\rm max}$-test, which provides an upper value of the discrepancy between the \Lm value expected by the concordance model and the alternative models here considered. We obtained that these future observations cannot discriminate the concordance model relative to
the KLCDM model. For example, $f_{max} \simeq 2$ at best for a Euclid-like  survey, assuming both realistic and optimistic configurations. By contrast, they can distinguish between LCDM and the dynamic dark energy models at a higher statistically significant level, i.e.,  $f_{max} \geq 5.3 \; (4.7)$ for  Euclid-like (SKA B1-like) configurations. Simulations assuming smaller uncertainties on $H(z)$ can reach $f_{max} \geq 6.6$

These results show that future redshift surveys are capable of falsifying the $\Lambda$CDM model without any {\it a priori} assumption of the nature of dark energy and cosmic expansion given that one of these three benchmark models truly describes the observed Universe.\\

\emph{Acknowledgments} -- 
CB acknowledges support from the Programa de Capacita\c{c}\~ao Institucional PCI/ON. MK acknowledge support from the Swiss National Science Foundation. RM acknowledges support from the South African Radio Astronomy Observatory (SARAO) and National Research Foundation (Grant No. 75415) support. CC and RM were supported by the UK Science \& Technology Facilities Council, Consolidated Grants ST/P000592/1 (CC) and ST/N000668/1 (RM). 

\end{document}